\documentclass[twocolumn]{aastex701}
\received{August 4, 2025}
\revised{August 25, 2025}
\accepted{August 30, 2025}
\begin{document}

\title{A carbon-rich disk surrounding a planetary-mass companion}

\author[0000-0001-7255-3251]{Gabriele Cugno}\altaffiliation{These authors contributed equally to this work}
\affiliation{Department of Astrophysics, University of Zurich, Winterthurerstrasse 190, 8057 Zürich, Switzerland}
\email[show]{gabriele.cugno@uzh.ch}  

\author[orcid=0000-0002-4022-4899]{Sierra L. Grant}\altaffiliation{These authors contributed equally to this work} 
\affiliation{Earth and Planets Laboratory, Carnegie Institution for Science, 5241 Broad Branch Road, NW, Washington, DC 20015, USA}
\email{sgrant@carnegiescience.edu}

\begin{abstract}
During the final assembly of gas giant planets, circumplanetary disks (CPDs) of gas and dust form due to the conservation of angular momentum, providing material to be accreted onto the planet and the ingredients for moons. The composition of these disks has remained elusive, as their faint nature and short separations from their host stars have limited our ability to access them. Now, with the spatial and spectral resolution of JWST/MIRI Medium-Resolution Spectrograph, we can observe and characterize this reservoir for wide-orbit planetary-mass companions for the first time. We present the mid-infrared spectrum from the CPD surrounding the young companion CT Cha b. The data show a carbon-rich chemistry with seven carbon-bearing molecules (up to C$_6$H$_6$) and one isotopologue detected and indicate a high gaseous C/O$>$1 that is in contrast with the elemental abundance ratios typically measured in directly imaged gas giant atmospheres. This carbon-rich chemistry is also in stark contrast to the spectrum of the disk surrounding the host star, CT Cha A, which shows no carbon-bearing molecules. This difference in disk chemistry between the host disk and its companion indicates rapid, divergent chemical evolution on $\sim$million-year timescales. Nonetheless, the chemical properties of the CPD follow trends observed in isolated objects, where disks transition from an oxygen-rich to carbon-rich composition with decreasing host mass. Our results provide the first direct insight into the chemical and physical properties of material being accreted onto a gas giant analogue and into its potential moon system.
\end{abstract}


\section{Introduction} 

Circumplanetary disks (CPDs) are an inevitable by-product of giant-planet formation, in the same way that protoplanetary disks are an outcome of star formation. They constitute the final reservoir of gas and solids available to giant planets while they are forming, regulate the accretion of such material, and are the birthplaces of satellite systems, the results of which we have in our own solar system (e.g., Jupiter’s Galilean moons; \citealt{LunineStevenson1982, CanupWard2002}). Characterizing CPD composition and evolution is thus essential to understand planet formation and, for the first time, provides an observational window into the moon formation process.

\begin{figure*}[ht!]
\includegraphics[width=\textwidth]{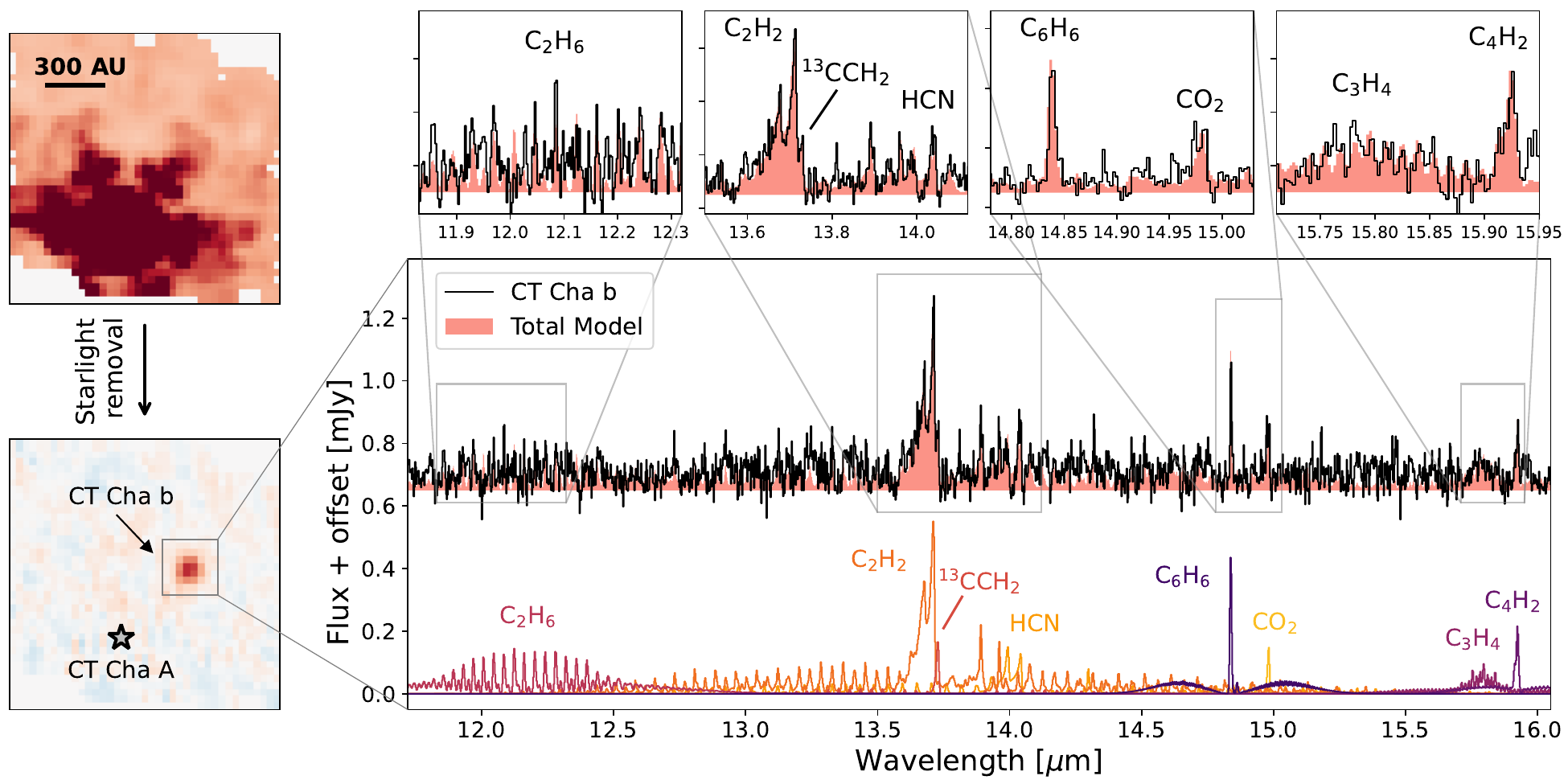}
\caption{Observed spectrum of CT Cha b. top-left: Calibrated JWST/MIRI MRS data of the CT Cha system, showcasing the stellar point-spread function dominating the image at $13.3-15.6~\mu$m. Bottom left: A spectral cross-correlation map of the same field of view revealing the companion. A gray star marks the location of the host star. Right: continuum-subtracted spectrum of CT Cha b (black) compared to a total model (red shaded area) composed of molecular emission from C$_2$H$_6$, C$_2$H$_2$, $^{13}$CCH$_2$, HCN, C$_6$H$_6$, CO$_2$, C$_3$H$_4$ and C$_4$H$_2$, shown with models in the colors below. The four panels on top show selected wavelength regions that contain important molecular features.
\label{fig:spectrum_b}}
\end{figure*}

Recent observational campaigns have revealed a handful of protoplanets and protoplanet candidates within circumstellar disks (e.g., \citealt{Keppler2018, Muller2018, Haffert2019, Hammond2023, Currie2022, Li2025}). Among them, PDS 70 c stands out as the only case thus far where radio emission, indicative of cold dust in the form of a CPD, has been detected \citep{Benisty2021}. Yet, despite such breakthroughs, direct detections of forming companions embedded in their protoplanetary disks remain extremely challenging. The obstacles are numerous: stringent contrast and angular resolution requirements, scattered light contamination from disk material \citep{Garufi2016}, and extinction along the line of sight (Cugno et al., subm.). 

To overcome these difficulties, the community has turned to more massive ($\times2-3$), farther out (few 100s au) counterparts as analogues. These planetary-mass companions span a broad range of properties, providing accessible laboratories for studying planet formation under diverse conditions. Accretion tracers were the first indication for the presence of disks around these companions, in systems like GQ Lup b, DH Tau b, and Delorme 1 (AB)b \citep[e.g.,][]{Seifahrt2007, Zhou2014, Eriksson2020, Demars2023}. Later, mid-infrared excesses \citep[e.g.,][]{MartinezKraus2022, Stolker2021} and sub-millimeter emission \citep{Wu2020} were detected for some of these sources, providing further evidence for the presence of a circumplanetary disk. The mid-infrared emission is particularly crucial, as at these wavelengths the grain size distribution, inner disk structure, and chemistry can be accessed, as is the case also for circumstellar disks. However, contrast and resolution limits, combined with poor atmospheric transmission at these wavelengths, have so far prevented their characterization. Recently, mid-infrared spectra of three CPDs were presented and analyzed \citep{Cugno2024, Patapis2025, Hoch2025}, demonstrating the improved capabilities of JWST. However, these spectra have low spectral resolution and limited spectral coverage, making the chemical composition inaccessible with those data.

\begin{figure*}[ht!]
\includegraphics[width=\textwidth]{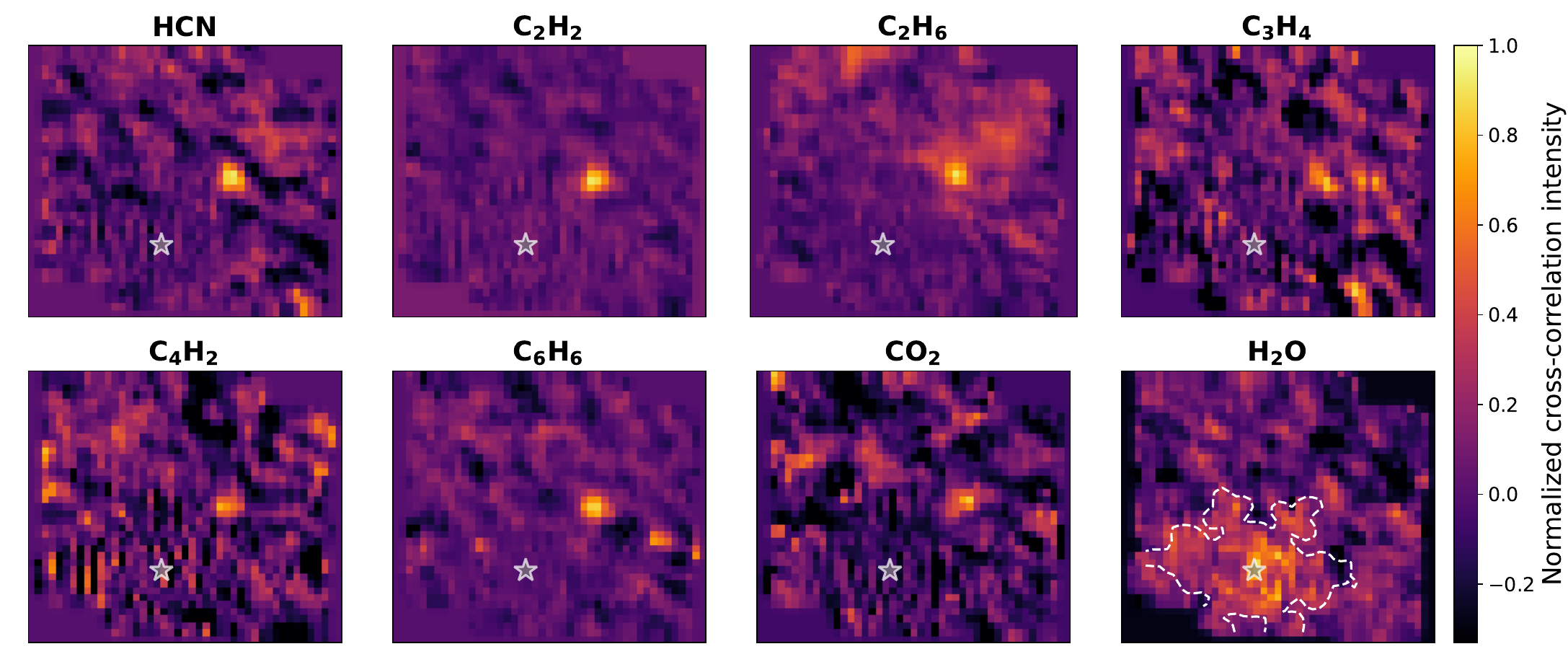}
\caption{Molecular SCC maps of the CT Cha system. For all of the molecules besides H$_2$O (bottom right), the SCC provides a signal at the location of CT Cha b, while no signal is present from CT Cha A, which is located in the lower left corner and indicated by the gray star. Conversely, the H$_2$O SCC is dominated by residual stellar features, and the shape of the stellar point-spread function from the original cube can still be recognized (white dashed line in the H$_2$O panel). 
\label{fig:SCC}}
\end{figure*}

In this Letter, we utilize the unprecedented angular and spectral resolution of JWST to provide the first detection and characterization of molecular emission from a CPD. We present JWST/MIRI Medium-Resolution Spectrograph (MRS) data of the CT Cha system, which consists of the host star CT Cha A (SpT K7, $M_* = 0.9\pm0.2~M_\odot$, \citealt{Weintraub1990, Ginski2024}) and the planetary-mass companion CT Cha b, a 14-24 M$_{Jup}$ \citep{Schmidt2008, Wu2015} object orbiting at a projected separation of 507 au. The companion has been found to be accreting at a rate of $6\times10^{-10}~M_\odot~\mathrm{yr}^{-1}$ via H$\alpha$ \citep{Wu2015} and Pa$\beta$ \citep{Schmidt2008}. These detections provide only very limited information on the accretion properties, but they suggest a reservoir of gas is present in the form of a CPD. However, until now emission from the disk itself has never been detected or characterized. In Sect.~\ref{sec:obs_dr} we introduce the observations and the data reduction. In Sect.~\ref{sec:results} we present the spectra of the disks around both CT Cha A and b, which are discussed and contextualized in Sect.~\ref{sec:discussion}. We summarize our findings and discuss the next steps in the study of CPD properties in Sect.~\ref{sec:conclusion}.

\section{Observations and Data reduction} \label{sec:obs_dr}

The CT Cha system was observed with the JWST/MIRI MRS as part of the General Observer (GO) program identifier (PID) 1958 (PI C. Rab) on August 15th, 2022. Fifty groups $\times$ six integrations were used in each of the three sub-bands (A, B and C) of every MRS channel (short, medium and long), yielding 56 minutes of on-source time per channel. The pointing was centered on CT Cha b, placing the host star outside the field of view in channels 1 and 2 and along the edge of the field in channels 3 and 4 where the pixel scale and field of view are larger. The observations span the full MRS wavelength range (4.9-27.9 $\mu$m) with a resolving power of R$\sim$3700–1300 and employed the four-point dither pattern optimized for point sources.

The raw {\tt uncal} files were retrieved from the Mikulski Archive for Space Telescopes (MAST) and calibrated with the JWST pipeline v1.18.0 under CRDS context 11.17.2 \citep{Bushouse2025}. Because the data were obtained during the first months of JWST operations, a hybrid reference-file strategy was adopted: Stage 1 processing used pmap 1118 to retain a contemporaneous bad-pixel mask, whereas Stages 2 and 3 applied pmap 1369, which provides the latest distortion solutions and flux-calibration files. Throughout the data reduction, we followed the standard Stage 2 recipe, including the default MRS fringe-correction routine. Spectral cubes were reconstructed in IFUALIGN mode, orienting each wavelength slice in the native MRS coordinate system and producing one two-dimensional image per wavelength element (see the top-left panel of Fig.~\ref{fig:spectrum_b}).

\begin{figure*}[ht!]
\includegraphics[width=\textwidth]{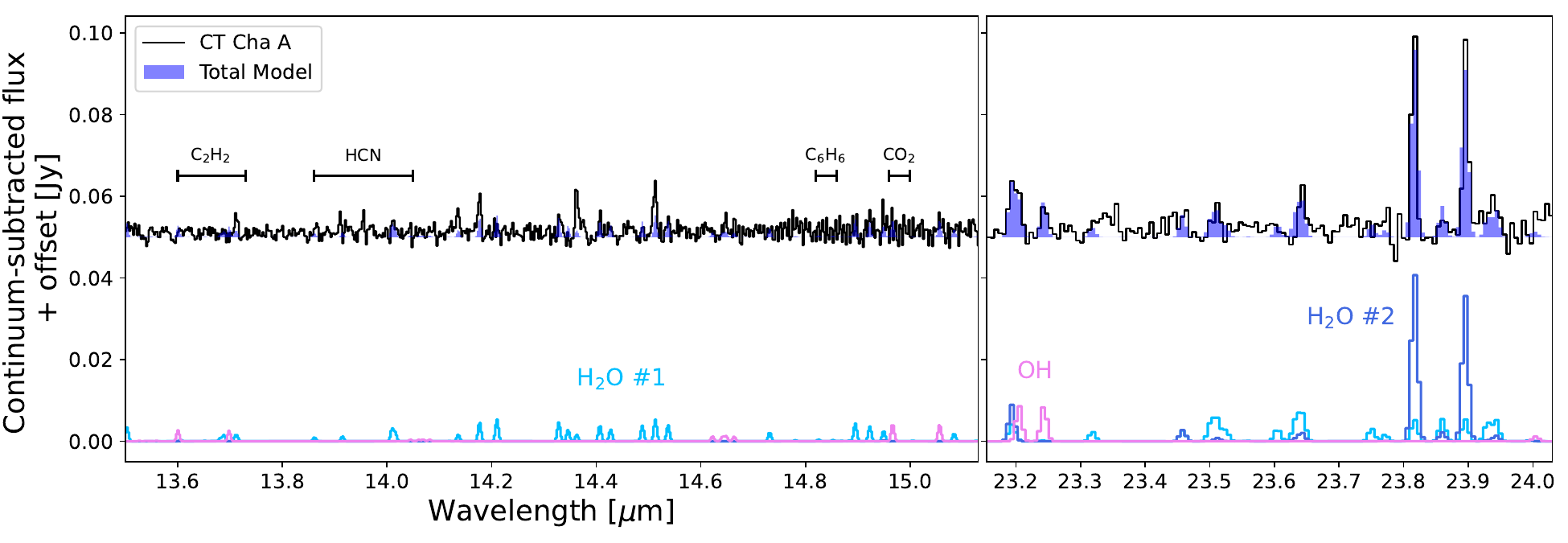}
\caption{Observed spectrum of CT Cha A. Left: CT Cha A (black line) extracted spectrum in the same spectral region in which the disk around the companion showcases emission features from carbon-bearing molecules, but where no such emission is detected in CT Cha A (regions indicated). The total best-fit model is shown in the blue shaded region with the individual molecular models from warm water ($T\sim$400 K, H$_2$O \#1), cold water ($T\sim$200 K, H$_2$O \#2), and OH shown below. Right: At longer wavelengths, bright lines from cold water dominate the spectrum. 
\label{fig:spectrum_A}}
\end{figure*}

At wavelengths longer than $10~\mu$m, the signal of the companion is embedded in the diffracted signal of the star, which is several orders of magnitude brighter (see top-left panel of Fig.~\ref{fig:spectrum_b}) and high-contrast spectroscopy techniques were required to detect and characterize emission from CT Cha b (see App.~\ref{app:extraction_b}). In particular, spectral cross-correlation (SCC) with an empirical template, namely the MRS spectrum of ISO-ChaI 147 from \cite{Arabhavi2024}, is used to identify the precise location of the companion. Given the robust detection of CT Cha b with SCC (see Fig.~\ref{fig:spectrum_b} and App.~\ref{app:extraction_b}), we tested whether it could be used to identify individual molecules as well. As SCC template spectra, we used the best-fit models from \cite{Grant2025} which reproduced the averaged spectra of nine very low-mass stars and brown dwarfs. For C$_2$H$_6$, which was not included in \cite{Grant2025}, we used best-fit model obtained from the fit of CT Cha b (see Sect.~\ref{sec:results}). We performed SCC following the procedure described in App.~\ref{app:extraction_b}. This approach revealed localized signals at the position of CT Cha b for C$_2$H$_6$, C$_2$H$_2$, HCN, C$_6$H$_6$, CO$_2$, C$_3$H$_4$, and C$_4$H$_2$ (Fig.~\ref{fig:SCC})\footnote{ The code used for post-processing and the final extracted spectra of this work are available at \url{https://github.com/gcugno/CTChab_extraction}}. Interestingly, cross correlating with the H$_2$O model shows strong contamination from CT Cha A (see bottom-right panel of Fig.~\ref{fig:SCC}). This is due to the presence of H$_2$O lines in the spectrum of A (see Sect.~\ref{sec:results}), and it reinforces the fact that a clear signal, as in the other panels, {can be obtained only when the companion exhibits distinct molecular features from the primary. This showcases that SCC is a powerful technique to identify CPD signals and characterize their content.}

Given the proximity of CT Cha b to its host (2.2 arcsec), the spectrum for the disk around CT Cha A can be extracted and compared to the disk around the companion. Due to the unusual position of the host star on the MRS detector, a customized extraction was required and only performed to extract the spectrum at wavelengths longer than $11.5~\mu$m (see App.~\ref{app:extraction_A}).

\section{Analysis and Results}\label{sec:results}

The spectrum of CT Cha b from 11.55 to $16.15~\mu$m is presented in the main panel of Fig.~\ref{fig:spectrum_b}. It contains emission features from multiple hydrocarbons (C$_2$H$_2$, C$_2$H$_6$, C$_3$H$_4$, C$_4$H$_2$, C$_6$H$_6$), HCN, and CO$_2$. The $^{13}$CCH$_2$ isotopologue is also detected. These robust detections indicate a very rich carbon chemistry in the CPD surrounding the companion. 

\subsection{Fitting the CT Cha b spectrum}
We reproduce the disk spectrum of CT Cha b using zero-dimensional, local thermodynamic equilibrium (LTE) slab models that provide constraints on the column density, temperature, and emitting area for each molecular species. Synthetic spectra are calculated following \cite{Tabone2023} and \cite{Arabhavi2024}, using molecular data (line positions, Einstein A coefficients, statistical weights, and partition functions) from the HITRAN \citep{Gordon2022} and GEISA \citep{Delahaye2021} databases. The model spectra are then calculated with three free parameters: the line-of-sight column density $N$, the gas temperature $T$, and the emitting area given by $\pi\,R^2$ for a disk of emission with radius $R$. We note that while we report the emitting area in terms of a radius, the emission could instead be coming from an annulus with the same equivalent area, so the radius should not be taken as a true disk radius. A grid of models is calculated for each molecular species with $N$ from $10^{14}$ to $10^{22}~\textrm{cm}^{-2}$, in steps of 0.166 in log$_{10}$-space, and $T$ from 100 to 1500 K, in steps of 25 K. The emitting area is varied by ranging the radius from 0.001 to 10 au in steps of 0.02 in log$_{10}$-space. Next, the model spectra are convolved to a resolving power of 2300 to be consistent with the data at these wavelengths, and then the model is sampled to the same wavelength grid as the data.  %The noise level is measured in a region which is as free from molecular emission as possible, which we find to be from 16.355 to $16.38~\mu$m in the CT Cha b spectrum, with a noise level of 0.019 mJy.  

We used the iterative model fitting approach from \cite{Grant2023}, wherein one molecular species is fitted, the best model is subtracted off, and another species is fit, continuing on through all of the detected molecules. To avoid contamination from other species, we have adopted the following fitting order, which goes from shortest to longest wavelength of the main features for each molecule: C$_2$H$_6$, C$_2$H$_2$, $^{13}$CCH$_2$, HCN, C$_6$H$_6$, CO$_2$, C$_3$H$_4$, then finally C$_4$H$_2$. Inspecting the residuals after subtracting the total model does not reveal additional molecular species with confidence. The best-fit model is shown in Fig.~\ref{fig:spectrum_b} (red shaded area) and the best-fit parameters are provided in Table~\ref{tab:best_fit}. Figure~\ref{fig:chi2b} shows the $\chi^2$ maps for each molecular species fit in the CT Cha b spectrum (see App.~\ref{app:chi2} for more details). The windows over which we fit each molecular species are provided in Table~\ref{tab:windows}. %The 1, 2, and 3$\sigma$ contours are calculated following \cite{Grant2023}.

\subsection{A carbon-rich disk surrounds CT Cha b}

We find that the C$_2$H$_2$ emission is the hottest gas species in the CT Cha b spectrum at $\sim$500 K and is just optically thick (see further discussion below). This is likely because we are probing the warm upper layers or close to the central object. The latter point is supported by the very small emitting area. C$_6$H$_6$, CO$_2$, and C$_4$H$_2$ are all fairly well constrained, with optically thin emission and cold temperatures ($\lesssim$250 K), similar to disks around isolated low-mass stars and brown dwarfs (e.g., \citealt{Arabhavi2024,Long2025, Grant2025}), and even carbon-rich disks around solar-type stars \citep{Colmenares2024}. Because these species are optically thin, there is a degeneracy between the column density and the emitting area, but the temperature is reliably cold. C$_2$H$_6$, HCN, and C$_3$H$_4$ are not well constrained, likely in part due to low signal-to-noise of the features and, in the case of HCN, contamination with other species. Additionally, as the line list for $^{13}$CCH$_2$ may not be complete, we do not draw strong conclusions from the fits of these species.

To estimate the carbon-to-oxygen ratio (C/O) of the CPD we followed \cite{Long2025}, who coupled the derived column density ratio $N_{\rm{C_2H_2}}$/$N_{\rm{CO_2}}$ with thermochemical models from \cite{Najita2011}. This method yields a C/O ratio for the disk surrounding CT Cha b of $\sim1.7$. This value should be taken with caution given that the presence of a pseudo-continuum is unclear (see below) and that the models from \cite{Najita2011} were constructed to study disks around T Tauri stars rather than disks around planetary-mass objects. Nonetheless, given the numerous other hydrocarbon species and the $^{13}$CCH$_2$ isotopologue detected, it is safe to assume that the disk around CT Cha b is very carbon-rich. 

One of the main constraints on the disk composition for the carbon-rich disks around low-mass stars and brown dwarfs is the C$_2$H$_2$ molecular pseudo-continuum, as it indicates very high abundances of carbon-rich gas \citep{Tabone2023, Arabhavi2024, Long2025}. Such a pseudo-continuum is not found in the CT Cha b spectrum, either because it is not present or due to the challenges in extracting the spectrum and removing the continuum contribution from CT Cha A. Therefore, only the main Q-branch can be fitted. Nonetheless, the best fit yields a C$_2$H$_2$ column density of $\sim3\times10^{17}$~cm$^{-2}$ (see Table~\ref{tab:best_fit}), a value that remains partially optically thick and should be regarded as a lower limit.

\subsection{Fitting the CT Cha A spectrum}
A direct comparison of the disks surrounding both a planetary-mass companion and its primary is possible only in a few known cases (e.g., CT~Cha, FU~Tau, SCH~J0359), as in many systems the disk around the host star is already dissipated. Despite this small sample, such analysis could provide useful insights as to how disks that likely formed from the same molecular cloud evolve and on what timescales. For this reason, in this section we present the spectrum of CT Cha A and its properties, which we compare with those of the disk around CT Cha b in Sect.~\ref{discussion_context}.

The final spectrum of CT Cha A in the $13.5-15.2~\mu$m and $23.15-24.05~\mu$m spectral ranges is presented in Fig.~\ref{fig:spectrum_A}. Although the spectrum is relatively noisy, given the observing set-up, no carbon-bearing species are detected. The lack of these species is reinforced by the SCC maps in Fig.~\ref{fig:SCC}, which show no signal from these molecules throughout the image that could be associated with the primary. Instead, the spectrum of CT Cha A only exhibits lines from H$_2$O and OH. Therefore, the disk around CT Cha A exhibits a very carbon-poor chemistry, in stark contrast to what is observed in CT Cha b.

Similarly to CT Cha b, we used zero-dimensional slab modeling to reproduce the spectrum for CT Cha A (Fig.~\ref{fig:spectrum_A}). First, we fit H$_2$O in regions of the spectrum dominated by warm water (H$_2$O \#1). This component is subtracted from the spectrum and then a model is fit in the region of low-energy lines (H$_2$O \#2) and subtracted. Finally, OH is fitted. The wavelength regions where the fit was performed for each component/species are provided in Table~\ref{tab:windows}. The best-fit model is shown in Fig.~\ref{fig:spectrum_A} and the best-fit parameters are provided in Table~\ref{tab:best_fit}. We note that OH is likely not in local thermal equilibrium \citep{Tabone2021} and therefore the best-fit parameters for OH should be taken with caution.

\begin{table}[h!]
\centering
\caption{Best-fit model parameters for CT Cha b and CT Cha A. We note that the molecular line list of $^{13}$CCH$_2$ is likely incomplete; therefore the provided values should be taken with caution.}
\begin{tabular}{lcccc}
\toprule
Molecule & $N$ [cm$^{-2}$] & $T$ [K] & $R$ [au] & Num. molecules \\
\hline
\multicolumn{5}{c}{CT Cha b} \\\hline
CO$_2$      & $1.50 \times 10^{14}$ & 100 & 8.7  & $7.80 \times 10^{42}$ \\
HCN         & $2.20 \times 10^{21}$ & 175 & 0.05 & $3.30 \times 10^{45}$ \\
C$_2$H$_2$  & $3.20 \times 10^{17}$ & 525 & 0.01 & $4.00 \times 10^{40}$ \\
$^{13}$CCH$_2$ & $6.80 \times 10^{14}$ & 150 & 0.68 & $2.20 \times 10^{41}$ \\
C$_2$H$_6$  & $2.20 \times 10^{18}$ & 200 & 0.05 & $3.60 \times 10^{42}$ \\
C$_3$H$_4$  & $1.50 \times 10^{18}$ & 175 & 0.04 & $1.40 \times 10^{42}$ \\
C$_4$H$_2$  & $1.50 \times 10^{16}$ & 100 & 0.36 & $1.30 \times 10^{42}$ \\
C$_6$H$_6$  & $6.80 \times 10^{16}$ & 125 & 0.30 & $4.20 \times 10^{42}$ \\
\hline\hline
\multicolumn{5}{c}{CT Cha A} \\\hline
H$_2$O \#1 & $2.20 \times 10^{19}$ & 425 & 0.17 & $4.60 \times 10^{44}$ \\
H$_2$O \#2 & $4.60 \times 10^{17}$ & 200 & 1.87 & $1.10 \times 10^{45}$ \\
OH         & $4.60 \times 10^{18}$ & 800 & 0.13 & $5.70 \times 10^{43}$ \\
\hline
\label{tab:best_fit}
\end{tabular}
\end{table}

\section{Discussion}\label{sec:discussion}

\subsection{The origin of the carbon-rich disk surrounding CT Cha b}

As presented in Sect.~\ref{sec:results}, the disk around the CT Cha b companion is very carbon-rich, and multiple hypotheses have been proposed to explain such carbon-rich inner disks. Efficient radial drift can carry oxygen-bearing ices rapidly inward, such that the oxygen-rich gas is accreted onto the central object very early, leaving a carbon-rich reservoir behind \citep{Mah2023, Long2025}. Alternatively, low dust opacities may allow these observations to probe deeper in the disk, where more carbon-rich gas may exist and at cooler temperatures, in line with the values derived for most of the molecules in the disk around CT Cha b \citep{Tabone2023, Arabhavi2025a, Grant2025}. Finally, the destruction of carbonaceous grains can further enhance the gas-phase carbon budget \citep{Tabone2023, Arabhavi2024, Colmenares2024}. If this carbon-rich gas is then transported outward \citep{Houge2025}, that could provide an alternative explanation for the cold temperatures derived for the hydrocarbon gas.

\begin{figure*}[ht!]
\centering
\includegraphics[scale=0.61]{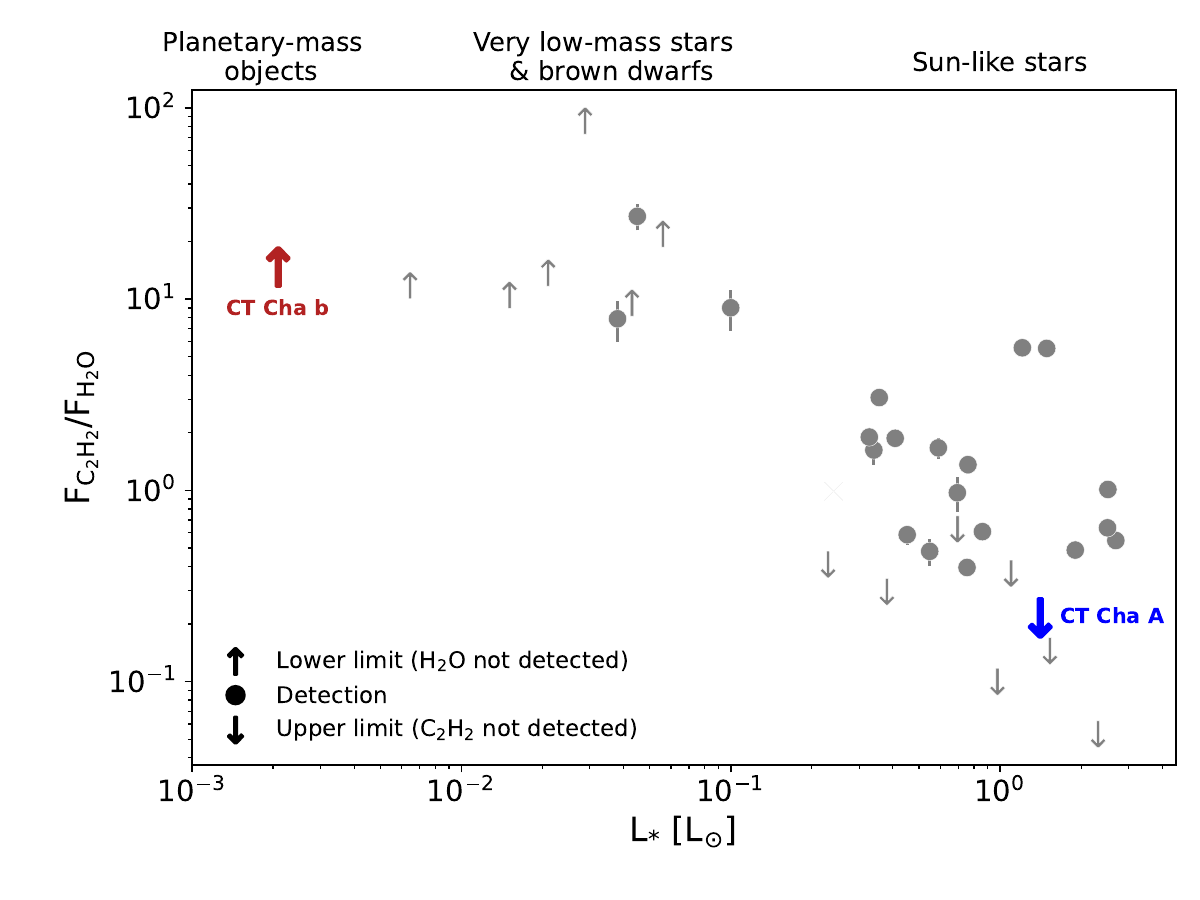}
\caption{The C$_2$H$_2$ to H$_2$O flux ratio as a function of host luminosity. The MINDS sample is shown in gray \citep{Grant2025}. CT Cha b is shown in red and CT Cha A in blue. Upward arrows are 3$\sigma$ lower limits (H$_2$O not detected) and downward arrows are 3$\sigma$ upper limits (C$_2$H$_2$ not detected).
\label{fig:isolated}}
\end{figure*}

While JWST/MIRI MRS data have been used to derive accretion luminosities \citep{Tofflemire2025}, there is no clear evidence of hydrogen recombination lines in the CT Cha b spectrum that we present here. The accretion rate for CT Cha b has been measured to be $\sim$6$\times$10$^{-10}$ M$_{\odot}$ yr$^{-1}$ \citep{Wu2015} and additional existing data (e.g., VLT/MUSE, VLT/UVES, JWST/NIRSpec) are more suited to explore accretion in this system in more detail and will be done so in coming works (e.g., Follette et al. in prep.). 
However, as CT Cha b is accreting from what we now know to be a carbon-rich CPD, it is worth exploring what this means for the atmospheric composition.
No C/O measurement is available in the literature for CT Cha b and deriving this value is beyond the scope of this work. However, medium- and high-resolution spectroscopic studies, combined with atmospheric retrieval analyses, have established robust atmospheric C/O ratios for several other wide-orbit ($\gtrsim10$~au) planetary-mass companions, with values clustering around 0.5–0.8 \citep[e.g.,][]{Hoch2023, Petrus2024, Nasedkin2024, Zhang2024, Gandhi2025, GonzalezPicos2025}. If CT Cha b follows the atmospheric C/O trends of these other companions, despite its carbon-rich disk, this would indicate that late-stage gas accretion is not able to substantially change the bulk C/O of the companion, especially in the higher end of the planet-mass regime. Therefore, the final phases of planet formation may have little impact on the atmospheric chemistry, and the planetary C/O ratio could trace its formation location and mechanism \citep{Oberg2011}.

\subsection{The cold-water-rich spectrum of CT Cha A}
\label{sec:discussion_ab}
The strongest features in the CT Cha A spectrum come from low-energy H$_2$O lines around $24~\mu$m coming from cold ($\sim200$~K) gas (H$_2$O \#2 in Fig.~\ref{fig:spectrum_A} and in Table~\ref{tab:best_fit}), consistent with sublimation of water ice \citep{Banzatti2023, Banzatti2025, Zhang2013}. A similar excess of cold water has been observed in GQ Lup \citep{Romero-Mirza2024}, another system that hosts a wide-orbit companion surrounded by a CPD \citep{Cugno2024}. Multiplicity can significantly impact dust evolution in disks, particularly by inducing efficient radial drift of dust grains \citep{Zagaria2021a,Zagaria2021b}, which can then transport ice-rich solids inward, enriching the inner disk gas with cold H$_2$O as the ice sublimates. Thus, in both GQ Lup and CT Cha, it is possible we are observing the chemical influence of wide-orbit companions on their host disks. While this is a tantalizing possibility, other explanations exist. For example, accretion variability, measured in CT Cha A \citep{HenizeMendoza1973, Carpenter2002, Kospal2012}, could cause the snowline to move outward, sublimating additional ices and enhancing the cold water vapor \citep{Houge_Krijt2023,Grant2024, Smith2025}. Another possibility is that cavities or warps of the inner disk relative to the outer disk could cause large areas of cold emission to be visible. Only with larger samples will we be able to determine if cold H$_2$O enrichment correlates with the presence of an outer companion or not.

%\subsection{The differing chemical evolution of CT Cha A and b}\label{sec:Ab_disks}

\subsection{Putting the CT Cha system into context}
\label{discussion_context}
If CT Cha b formed in situ, it likely formed via cloud collapse, as not enough disk material is available at several hundred astronomical units to form such a massive companion. Hence, CT Cha A and b formed from the same cloud material and at the same time. Therefore, by comparing the gas composition and properties of the two objects we can learn about the chemical and physical evolution and processing of material in each disk. The detected molecular species showcase a wholly different composition: 
CT Cha b hosts a very carbon-rich disk, with the oxygen coming only in the form of CO$_2$, while 
its primary contains only oxygen-bearing species, with no carbon-bearing molecules detected. This disparate chemistry implies that efficient local processes to each disk are driving the chemical evolution, producing these significant differences in only 1.6 Myr, the age of the system \citep{Galli2021}. 

Such disparate chemistry may not be surprising when we look to isolated objects for comparison. Changes in the observed disk composition as a function of stellar mass/luminosity have long been known, as Spitzer showed that disks around stars with spectral types later than M5 had stronger C$_2$H$_2$ emission relative to HCN, while earlier spectral types had the opposite \citep{Pascucci2009}. With JWST, the chemical differences are increasingly stark. There is a strong anticorrelation between the flux ratio of C$_2$H$_2$ to H$_2$O as a function of stellar luminosity \citep{Grant2025} that was found recently using data from the Mid-INfrared Disk Survey (MINDS) JWST Guaranteed Time program (PID 1282, PI T. Henning, \citealt{Henning2024, Kamp2023}). We show this correlation in Figure~\ref{fig:isolated}, now adding in CT Cha A and b. While many disks around Sun-like stars have some carbon-bearing species, it is not universal \citep{Pontoppidan2010, Arulanantham2025} and CT Cha A, with its nondetection of C$_2$H$_2$, is well in line with other objects with similar stellar luminosities. At the other end of the luminosity range, CT Cha b, with its carbon-rich chemistry, is similar to other disks around isolated low-mass stars, brown dwarfs, and free-floating planetary-mass objects \citep{Tabone2023, Arabhavi2024,Arabhavi2025b,Kanwar2024, Long2025, Flagg2025}, but now seen in a disk around a planetary-mass companion. This indicates that whatever process is driving the carbon-rich chemistry in low-mass stars and brown dwarfs extends down to the planetary-mass regime and in disks around wide-orbit companions.

\section{Summary and conclusions}\label{sec:conclusion}

We present the first medium-resolution, mid-infrared spectrum of a disk around a planetary-mass companion, CT Cha b. 

\begin{itemize}
    \item We used three-dimensional spectral cross-correlation to detect the CPD around the forming companion, demonstrating that this is an effective technique to identify companions and characterize their disks. With the planet location determined, we extracted its spectrum. 

    \item We identify C$_2$H$_6$, C$_2$H$_2$, HCN, C$_6$H$_6$, CO$_2$, C$_3$H$_4$, C$_4$H$_2$, and the isotopologue $^{13}$CCH$_2$ in the emission spectrum from the circumplanetary disk around CT Cha b. 

    \item We used LTE slab models to reproduce the spectrum, providing constraints on the gas properties for each molecular species. The column density ratio of C$_2$H$_2$ to CO$_2$, compared with thermochemical models, suggests C/O$>$1. 
    
    \item We find no carbon-bearing species in the spectrum from the disk around the host star, and we identify only H$_2$O and OH. The most prominent features in the CT Cha A spectrum are low-energy H$_2$O lines coming from cold ($\lesssim$200 K) water. 

    \item Comparing CT Cha b and A to isolated disks with a range of host properties, we find that CT Cha A's disk chemistry is well in line with other T Tauri objects at similar stellar luminosities. CT Cha b, on the other hand, extends line flux correlations and observations of carbon-rich chemistry in disks around low-mass objects down to a previously unexplored parameter space.  
\end{itemize}

Disks around planetary-mass companions, such as the disk around CT Cha b analyzed here, represent a new frontier in our understanding of planet formation and the first opportunity to observationally characterize moon-forming environments. Understanding the chemical and physical evolution occurring in CPDs is key to shedding light on the striking compositional diversity of solar system moons, best exemplified by the water-dominated Europa \citep{Kivelson2000} and the carbon-rich Titan \citep{Niemann2005}. 

Given the unprecedented sensitivity and angular resolution of JWST, more CPDs can be observed and studied, and the census is expected to expand as additional disks are discovered around directly imaged, still-forming planets \citep[e.g.,][]{Hoch2025, Patapis2025}. In the coming year, all of the currently known CPDs accessible to JWST/MIRI MRS will have been observed, providing a sample of nine sources, large enough for a study at the population level. This dataset will let us compare CT Cha b with companions of similar spectral type, mass, and luminosity, and search for systematic trends with global properties such as system age and planet–star separation.

\begin{acknowledgments}
The authors thank the referee for a constructive report that helped improve the manuscript. The authors thank Christian Rab and the proposing team for leading the program to obtain this dataset. The authors thank Beno\^it Tabone for developing the code used to fit the spectra. The authors thank Christian Rab, Polychronis Patapis, Helena Kühnle, Valentin Christiaens, Elena Kokoulina, and Olivier Absil for useful discussions. GC thanks the Swiss National Science Foundation for financial support under grant number P5R5PT\_225479. This work is based on observations made with the NASA/ESA/CSA JWST. The data were obtained from the Mikulski Archive for Space Telescopes at the Space Telescope Science Institute, which is operated by the Association of Universities for Research in Astronomy, Inc., under NASA contract NAS 5-03127 for JWST. These observations are associated with program JWST-GO-01958 and can be accessed via DOI: \url{http://dx.doi.org/10.17909/1v33-tr24}. The code used to perform the spectral cross-correlation and extract the spectrum of CT Cha b can be found at \url{https://github.com/gcugno/CTChab_extraction} or \url{https://zenodo.org/records/16944199} \citep{Cugno_codes}, in addition to all the data used in this Letter.
\end{acknowledgments}

\begin{contribution}

G.C. and S.G. carried out the data reduction, analysis, and interpretation. G.C. identified the location of the companion and extracted and cleaned the spectra of both the companion and the primary. G.C. performed the spectral cross correlation. S.G. provided the molecular templates for the cross correlation and performed the spectrum fitting for both the primary and the companion. S.G. provided the data for Figure~\ref{fig:isolated}. G.C. and S.G. contributed equally to the writing of the manuscript.

\end{contribution}

\facilities{JWST(MIRI)}

\software{jwst version 1.18.0 \citep{Bushouse2025}}


\appendix

\section{Spectral extraction of CT Cha b}\label{app:extraction_b}

To pinpoint the companion signal in the MRS data cubes we applied a three-dimensional spectral-cross-correlation (SCC) technique widely used in high-resolution spectroscopic studies of exoplanets \citep{Hoeijmakers2018} and protoplanets \citep{Cugno2021}, and predicted to work on MRS data \citep{Patapis2022, Malin2023}. SCC quantifies the similarity between an observed spectrum and a template on a spaxel-by-spaxel basis to isolate localized spectral features in the datacube. A peak in the resulting maps indicates the presence of the template’s line pattern in the data. As a template, we adopted the spectrum of a disk around a very low-mass star, ISO-ChaI 147 \citep{Arabhavi2024}. Its spectrum exhibits prominent emission features characteristic of objects with masses and spectral types comparable to CT Cha b, in particular dominated by features arising from a multitude of hydrocarbons. 

Starting from the fully calibrated 3D cubes, we mitigated large-scale background and stellar contamination by subtracting, from each spaxel, its own spectrum smoothed with a Gaussian kernel ($\sigma$ = 10 pixel; \citealt{Patapis2022}). The same high-pass filter was applied to the template spectrum. After normalizing each spaxel to its peak value, we cross correlated every spaxel spectrum with the template, yielding an SCC map such as the one shown in the lower-left panel of Fig.~\ref{fig:spectrum_b}. The map reveals a single, unresolved signal matching the template line pattern in the $11.5-16.1~\mu$m region. The strong correlation between the data and the template indicates that the emission from the disk around CT Cha b arises from a mixture of hydrocarbons. If the spectrum of the disk around the host star were equally rich in the same molecules, no isolated SCC signal would emerge (see Sect.~\ref{sec:results}), underscoring that the detected signal is unique to CT Cha b.

\begin{figure*}[ht!]
\includegraphics[width=\textwidth]{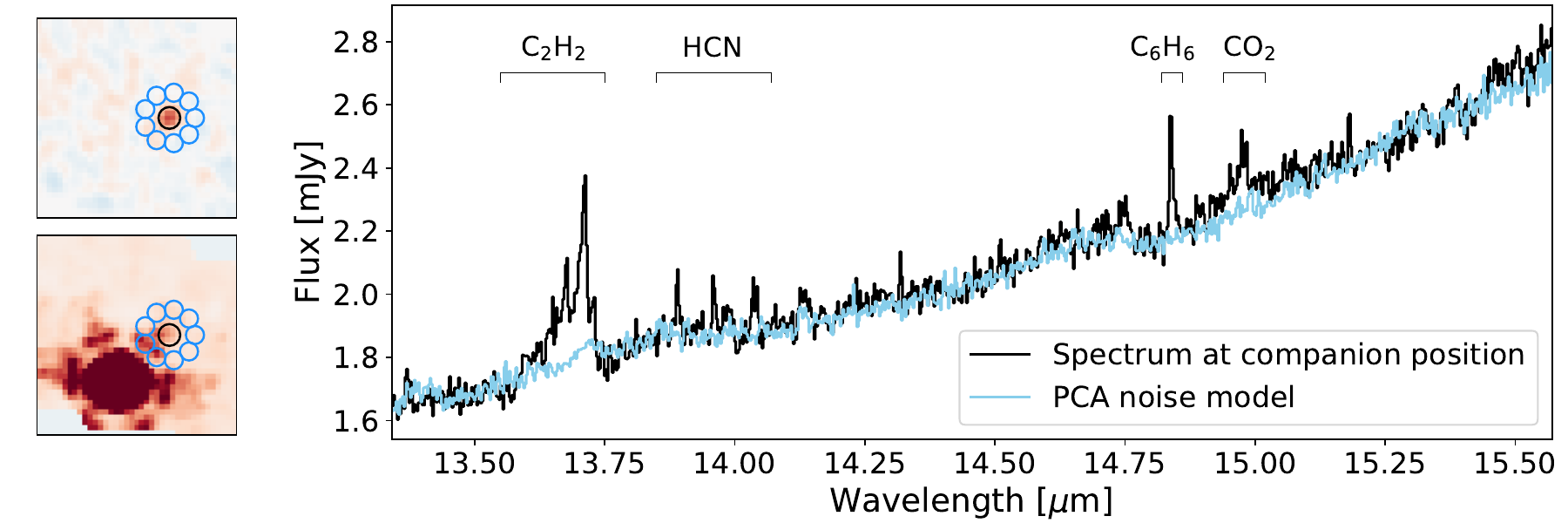}
\caption{Extraction process of the CT Cha b spectrum. Left: 2D SCC map (top) showing the signal of CT Cha b (black) and the neighboring apertures (blue) used to sample the instrumental and stellar noise from the calibrated data (bottom). Right: The spectrum extracted at the location of the companion is shown in black, while the projection of the PCA basis obtained from the noise apertures is shown in blue. This spectrum encapsulates various noise sources, like the signal from the stellar point-spread function, the background, and instrumental and detector effects.
\label{fig:noise_rem_b}}
\end{figure*}

We fitted a two-dimensional Gaussian to each SCC map obtained with ISO-ChaI 147 to determine the CPD position and then utilized the {\tt extract1d} step of the {\tt jwst} reduction pipeline to recover the spectrum at that location, using a circular aperture of radius 0.5 $\times$ FWHM to minimize contamination from CT Cha A. This function applies an aperture correction. Spectra in nine identical apertures, offset by five spaxels relative to the CPD (see the left panels of Fig.~\ref{fig:noise_rem_b}), were extracted in an identical fashion to characterize the local noise budget and behavior, capturing both spectral features due to residual stellar light and instrumental background. These off-source spectra were assembled into an orthogonal basis via principal-component analysis (PCA); the first five components were projected onto the CPD spectrum and subtracted (see the right panel of Fig.~\ref{fig:noise_rem_b}). Finally, residual low-frequency fluctuations remained in the extracted spectrum, possibly due to varying wavelength-dependent diffraction from the stellar point-spread function at the location of the companion. To remove residual continuum contributions, we applied continuum subtraction using the Iterative Reweighted Spline Quantile Regression algorithm of \texttt{pybaselines} \citep{Erb2022} following \cite{Temmink2024} and masking the region between 13.55 and $14.4~\mu$m to preserve the potential pseudo-continuum from C$_2$H$_2$ \citep{Tabone2023, Arabhavi2024, Long2025}. The procedure removes the persistent baseline variations and yields the final spectrum of CT Cha b which is presented in Figure~\ref{fig:spectrum_b}.

\section{Spectral extraction of CT Cha A}\label{app:extraction_A}
The spectrum of CT Cha A was only obtained for channels 3A–4C ($11.5-27~\mu$m), as at shorter wavelengths the stellar photocenter did not fall in the field of view of the data cube and could not be extracted. For each cube the location of the star was identified in the wavelength-collapsed image with a 2D Gaussian, and the resulting coordinates were passed to the {\tt extract1d} step of the {\tt jwst} pipeline, which extracted the spectrum in a circular aperture with a radius 1.5 $\times$ FWHM, yielding a one-dimensional spectrum for each channel. Due to the unusual location of the star on the detector, the extracted spectra suffered from strong systematics, reminiscent of fringes, and a large number of outliers. Fringes significantly depend on the location of the object on the detector \citep{Argyriou2020}, and a suboptimal correction due to the extreme location of CT Cha A compared to most other MRS datasets is likely. We verified that similar features are present in another dataset whose primary star is located in a similar position on the detector (PID 3647, PI P. Patapis), confirming the instrumental nature of these systematics.

\begin{figure*}[ht!]
\includegraphics[width=\textwidth]{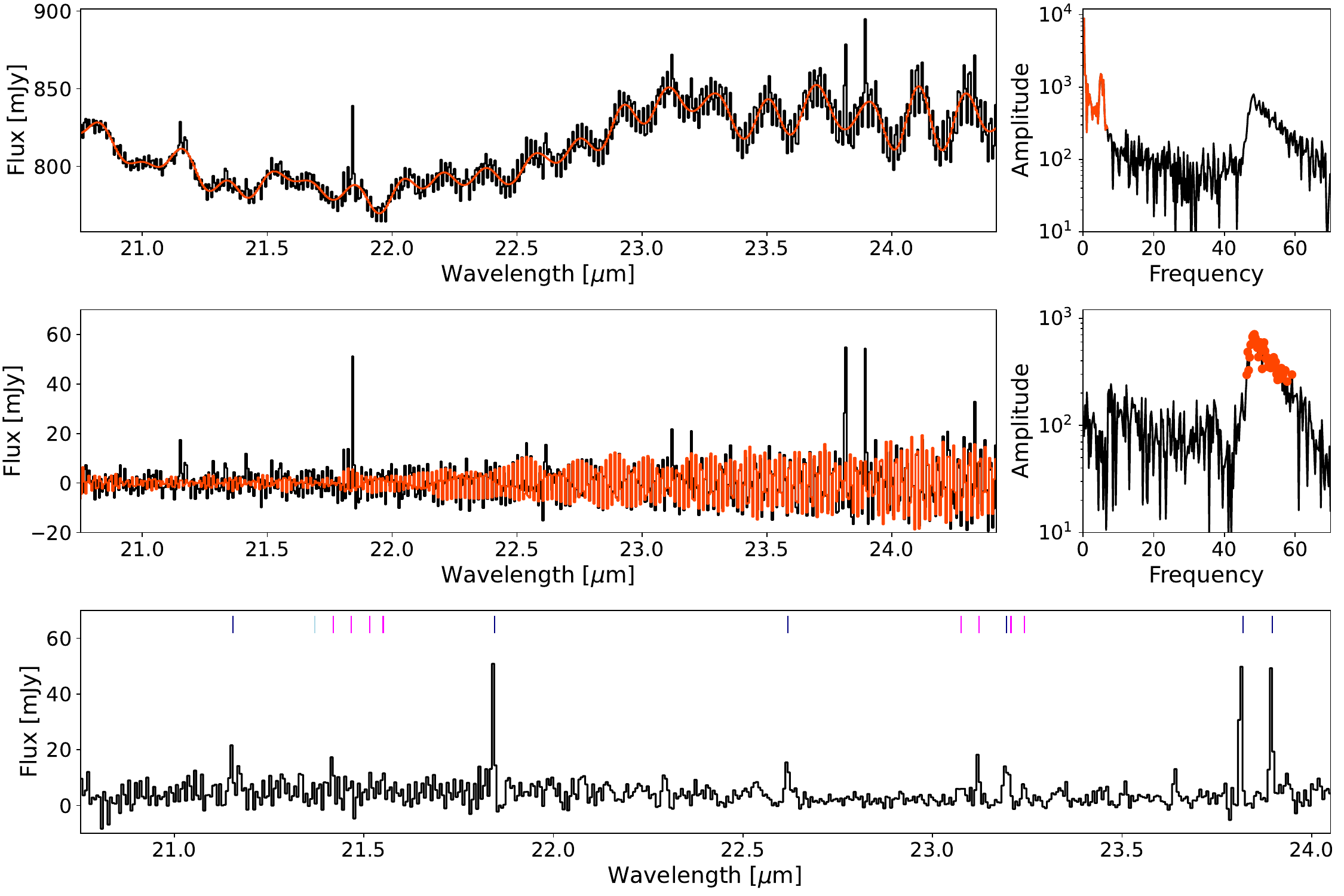}
\caption{Postprocessing of the spectrum of CT Cha A. Top row: Original extracted 1D spectrum from the star (black, left panel) and the corresponding modes (right panel). The modes highlighted in orange have been used to reconstruct the low-frequency modulation of the spectrum, shown in orange in the left panel. Middle row: Same as the top panels, but for the modes with the highest amplitudes. Bottom row: Final continuum-subtracted spectrum of CT Cha A in channel 4B. Vertical lines at the top of the panel indicate the location of expected H$_2$O (blue) and OH (pink) emission lines. Our method is able to retain most of the expected emission lines in this range, distinguishing them from the systematic noise initially present. 
\label{fig:noise_rem_A}}
\end{figure*}

To remove these instrumental systematics, each extracted sub-band spectrum underwent a multistage cleaning sequence. Low-frequency modulations were modeled in Fourier space: components below a band-specific cutoff frequency were isolated, inverse-transformed, and subtracted from the data (see top row of Fig.~\ref{fig:noise_rem_A} for channel 4B, $21.7-24.3~\mu$m). For this, a version of the spectrum was used in which emission lines identified through a sigma-clipping routine (2.5 $\sigma$) were substituted with the local continuum estimate. Any remaining high-frequency instrumental noise was suppressed by removing a channel-specific number of the strongest Fourier modes (varying band by band, between 30 and 60) and subtracting the reconstructed noise signal (see middle row of Fig.~\ref{fig:noise_rem_A}  for channel 4B). This method removes low- and high-frequency components, preserving only astrophysical features (bottom panel of Fig.~\ref{fig:noise_rem_A}).

\section{$\chi^2$ maps}\label{app:chi2}
\begin{figure*}[ht!]
\includegraphics[width=\textwidth]{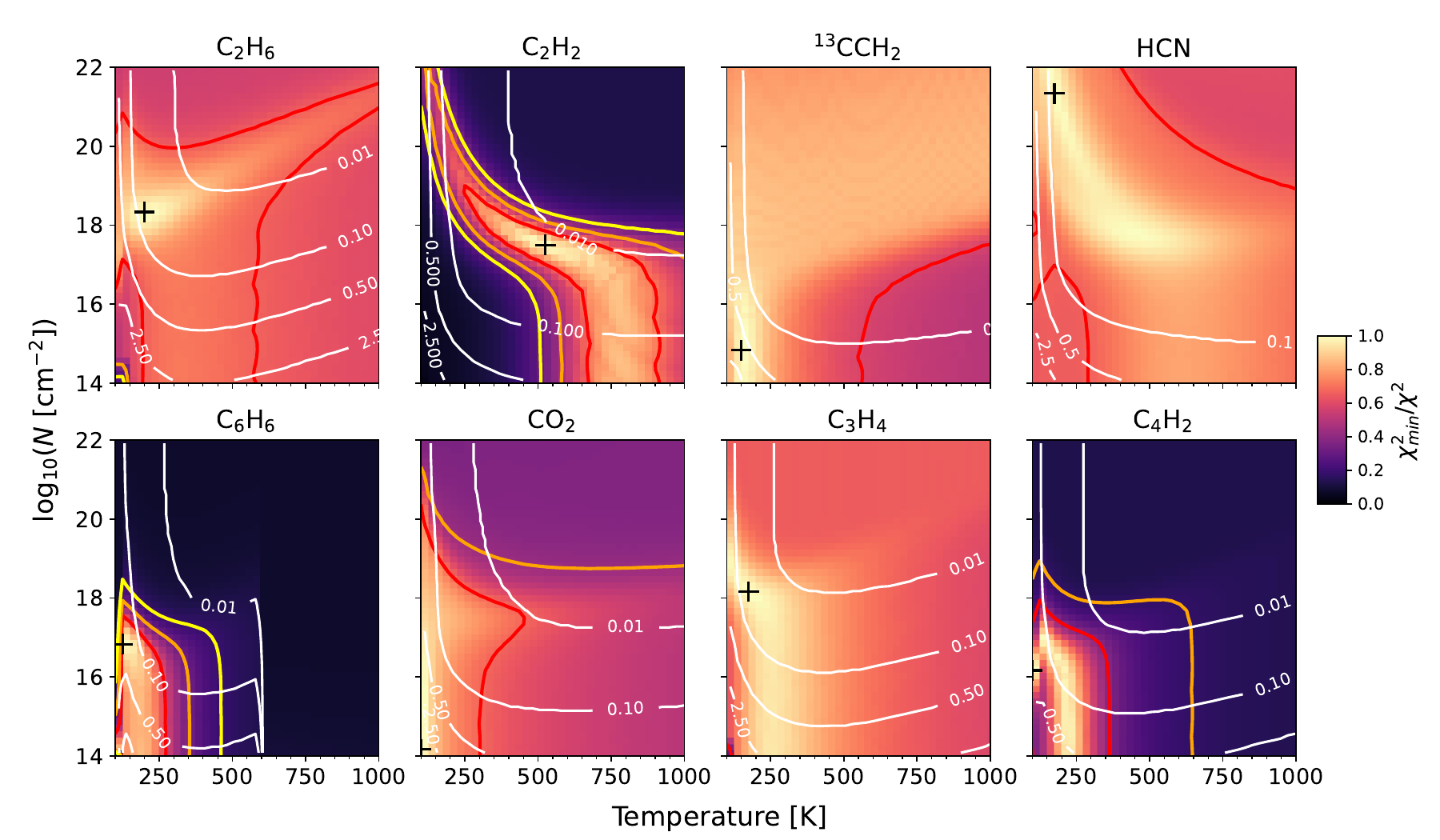}
\caption{$\chi^2$ maps for the molecules detected in the spectrum of CT Cha b. The model corresponding to the best-fit is represented by the black plus. The 1, 2, and 3$\sigma$ contours are shown in yellow, orange, and red, respectively. Equivalent emitting radii are shown in the labeled white contours with the radii in au. The explored parameter space goes up to temperatures of 1500 K, but we show a zoomed-in temperature range for clarity.
\label{fig:chi2b}}
\end{figure*}

\begin{figure*}[ht!]
\includegraphics[width=\textwidth]{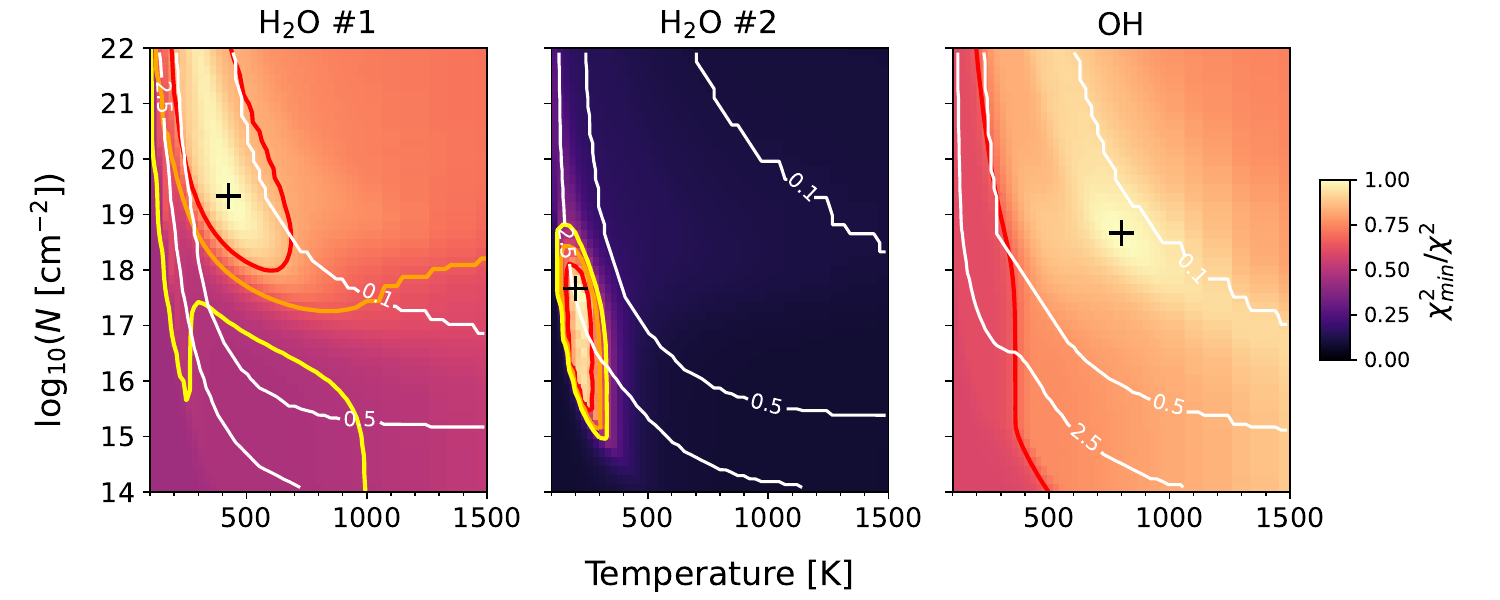}
\caption{Same as Figure~\ref{fig:chi2b}, but for the molecules identified in the CT Cha A spectrum. 
\label{fig:chi2A}}
\end{figure*}

The reduced $\chi^2$ maps for the molecules analyzed in the spectra of CT Cha b and CT Cha A are presented in Figures~\ref{fig:chi2b} and~\ref{fig:chi2A}, respectively. The maps are calculated following \cite{Grant2023}. The reduced $\chi^{2}$ is determined using the following formula: 
\begin{equation}
    \chi^2 = \frac{1}{N}\sum_{i=1}^{N} \frac{(F_{obs, i}-F_{mod, i})^2}{\sigma^2},
\end{equation}
where $N$ is the number of resolution elements in the spectral windows that the fit is done over and $\sigma$ is the standard deviation. For CT Cha A and b in the fit to the $\sim$12 to 16 $\mu$m region, we calculate the noise in the window from 16.355 to 16.38 $\mu$m, resulting in 0.019 mJy for CT Cha b and 0.761 mJy for CT Cha A. For the OH and second H$_2$O component in CT Cha A which are fitted at the longer wavelengths, we measure the noise from 22.7 to 22.85 $\mu$m, resulting in a noise level of 1.286 mJy. We note that the emitting area is simply a scaling factor, while the column density and temperature control the spectral shape and the relative strength of the lines. Therefore, for each column density and temperature, the equivalent emitting radius is determined from scaling the model flux to match the observed spectrum, and $N$ and $T$ are the only free parameters. The contours in the reduced $\chi^2$ shown in the figures are the 1$\sigma$, 2$\sigma$, and 3$\sigma$ levels determined as $\chi^2_{min}$ + 2.3, $\chi^2_{min}$ + 6.2, and $\chi^2_{min}$ + 11.8, respectively (see \citealt{numericalrecipesinc} and Table 1 and Eq. 6 of \citealt{Avni1976}). The wavelength windows that we use to fit each molecular species are provided in Table~\ref{tab:windows}.

\begin{table}[h!]
\centering
\caption{The windows over which each molecule is fit for CT Cha b and CT Cha A}
\begin{tabular}{lc}
\toprule
Molecule & Fitting range [$\mu$m] \\
\hline
\multicolumn{2}{c}{CT Cha b} \\\hline
CO$_2$      & 14.977--14.987, 16.18--16.2  \\
HCN         &  13.91--14.05 \\
C$_2$H$_2$  & 13.61--13.72 \\
$^{13}$CCH$_2$ & 13.72--13.74 \\
C$_2$H$_6$  & 12.075--12.092, 12.155--12.17, 12.23--12.25,12.39--12.41 \\
C$_3$H$_4$  & 15.72 --15.87 \\
C$_4$H$_2$  & 15.9--15.94 \\
C$_6$H$_6$  & 14.835--14.875, 15.04--15.06 \\
\hline\hline
\multicolumn{2}{c}{CT Cha A} \\\hline
H$_2$O \#1 & 14.47--14.55, 16.1--16.125, 17.2--17.25, 22.9--22.94 \\
H$_2$O \#2 & 22.6--22.65, 23.8--23.95 \\
OH         & 13.689--13.707, 14.036--14.09, 14.61--14.68, 14.95--15.1, 15.89--16.06, 19.95--20.2, 21.25--21.6, 22.95--23.15 \\
\hline
\label{tab:windows}
\end{tabular}
\end{table}

%% For this sample we use BibTeX plus aasjournalv7.bst to generate the

\bibliography{main}{}
\bibliographystyle{aasjournalv7}



\end{document}